\documentstyle[preprint,eqsecnum,aps]{revtex}

\begin{document}
\title{Bouncing open universes embeddable in a distorted Randall-Sundrum brane
scenario}
\author{Rolando Cardenas\thanks{%
rcardenas@uclv.etecsa.cu}, I. Quiros\thanks{%
israel@uclv.etecsa.cu} and Rolando Bonal\thanks{%
bonal@uclv.etecsa.cu}}
\address{Dpto. Fisica. Universidad Central de Las Villas. Santa Clara CP 54830. Villa%
\\
Clara. Cuba}
\date{\today}
\maketitle

\begin{abstract}
In reference \cite{qbc} a four-dimensional effective theory of gravity
embeddable in a five-dimensional "distorted" Randall-Sundrum brane scenario
was derived. The present paper is aimed at the application of such a theory
to describe physics in an open Friedmann-Robertson-Walker (Weyl-symmetric)
universe. It is shown that regular bouncing universes arise for a given
range of the free parameter of the theory.
\end{abstract}

\bigskip

One of the main ingredients that enter the basis of the unification scheme of the fundamental interactions is that multiple dimensions are required. In fact, modern field theory suggests spacetime to be $4+n$-dimensional ($n=6$ for string theory and $n=7$ for supergravity). It is usually assumed that the additional dimensions may be compacitifed down to size of the order the Planck length or below so they are not observable. However, recent developments suggest that some of the additional dimensions may have compactification radii larger than the Planck length without conflict with observations\cite{arkani-hamed}. Besides, it is possible that gravity remains confined to a four-dimensional slice of the bulk spacetime leading to extra-dimensions being even infinite in extent\cite{randall}. 

In reference\cite{qbc} a four-dimensional effective theory of gravity embeddable in a five-dimensional "distorted" Randall-Sundrum brane scenario was derived. The fifth extra dimension was identified with the dilaton field $\psi$. We started with the Randall-Sundrum scenario improved in Ref.\cite{lykken} and then we "distorted" it by relaxing some requirements as, for instance, orbifold symmetry and Poincare invariance. Then we elaborated on this "distorted" five-dimensional brane scenario by studying the effective geometry induced on the four-dimensional manifold. It resulted in a Weyl-integrable geometry that may be seen as a distorted Riemann geometry. Then we postulated a four-dimensional theory of gravity that could be embedded in the higher-symmetric structure, i.e., a theory sharing some of the basic properties of the "distorted" brane set-up worked out previously: the theory  should be built over Weyl-integrable geometry (then it should be Weyl-symmetric), the matter degrees of freedom should be coupled to the metric induced on the "visible" brane of the higher-dimensional scenario instead of the four-dimensional "Planck" metric, there should be some "signal" from the extra-dimension, etc. Respecting the cosmological issue, the singularity problem was treated for the case of flat Friedmann-Robertson-Walker universes. Now we apply this theory for the description of physics in an open Friedmann-Robertson-Walker (Weyl-symmetric)
universe.  

We start with an action similar to that used in \cite{qbc}:

$$
S=\int d^4x\sqrt{-g}e^\psi (R-\omega (\nabla \psi )^2+16\pi e^\psi {\cal L}%
_m), \eqno{(1)}
$$
where $\psi$ is the dilaton field (linked with the extra dimension), $R$ is
the curvature scalar of the metric $g_{ab}$, $\omega$ \ is a free parameter%
\footnotemark 
\footnotetext{
It is related with the free parameter in Ref.\cite{qbc} through $\omega=\frac{1}{4k^2}-\frac{3}{2}$} and ${\cal L}_m$ is the Lagrangian of the ordinary
matter fields. The change of the sign before $\psi$ in the exponents, with
respect to Ref.\cite{qbc}, does not affect the physics. The variational
principle gives

\bigskip 
$$
R_{ab}-\frac 12g_{ab}R =8\pi e^{-\psi }T_{ab}+\omega [\psi _{,a}\psi _{,b}-
\frac 12g_{ab}(\nabla \psi )^2]+[-\psi _{||ab}+\psi _{,a}\psi _{,b}+g_{ab}\Box \psi -g_{ab}(\nabla \psi
)^2], 
\eqno{(2)}
$$

$$
(\nabla \psi )^2+\Box \psi =0 
\eqno{(3)}
$$
where $T_{ab}=\frac 2{\sqrt{-g}}\frac{\delta (\sqrt{-g}{\cal L}_m)}{ \delta
g_{ab}}$, $\Box \psi =g^{mn}\psi _{||mn}$, and the bar denotes covariant
differentiation in a metric sense, i.e., defined through the Christoffel
symbols of the metric $g_{ab}$. Recall that the theory is built over
Weyl-integrable geometry so the affine connection does not coincide with the
Christoffel symbols. From this equation one sees that Newton's constant $%
G=e^{-\psi }$.

For an open Friedmann-Robertson-Walker(FRW) universe filled with a
barotropic perfect fluid, the above field equations can be written
explicitly in the following way

$$
(\frac{\stackrel{.}{a}}a)^2-\frac 1{a^2}=-\frac M{e^{(\frac{3\gamma }2%
+1)\psi }a^{3\gamma }}+\frac{\omega \stackrel{.}{\psi ^2}}6+\frac{\stackrel{.%
}{a}}a\stackrel{.}{\psi },  
\eqno{(4)}
$$

$$
\ddot{\psi}+3\frac{\stackrel{.}{a}}a\stackrel{.}{\psi }+\stackrel{.}{\psi }%
^2=0
\eqno{(5)}
$$

$M$ is an arbitrary integration constant and the barotropic index $\gamma $
is in the range $0<\gamma <2$. While deriving eq.(4) we have taken into
account that the ordinary matter energy density is given by $\mu =\frac 3{%
8\pi }\frac M{e^{\frac{3\gamma \psi }2}a^{3\gamma }}$. After integrating
once the wave equation for the scalar field $\psi $, we obtain:

$$
(e^\psi )^{.}=\pm \frac{\sqrt{N}}{2a^3}. 
\eqno{(6)}
$$

where N is another integration constant.

\vspace{0.2in}Substitution of eq. (6) into Eq. (4) gives:\mathstrut
\mathstrut 
$$
\stackrel{.}{a}^2-1=-\frac M{e^{(\frac{3\gamma }2+1)\psi }a^{3\gamma -2}}+%
\frac{N\omega }{24e^{2\psi }a^4}\pm \frac{\sqrt{N}\stackrel{.}{a}}{2e^\psi
a^2},  
\eqno{(7)}
$$

The curvature scalar for an open FRW universe is found to be

$$
R=\frac{3M(4-3\gamma )}{e^{(\frac{3\gamma }2+1)\psi }a^{3\gamma }}-\frac{
N\omega }{4e^{2\psi }a^6}  
\eqno{(8)}
$$

The high complexity of the system of equations (6)-(7) and of eq. (8) leads
us to the use of the conformal transformation technique\cite{far}. Well worth to notice that the conformal theory will be Einstein's canonical
general relativity.

We shall use the Raychaudhuri equation\cite{haw}. For a congruence of fluid lines
without vorticity and shear, with the time-like tangent vector $\hat k%
^a=\delta _0^a$, it can be written in the conformal frame as:

$$
\dot{\hat\theta }=-\dot{\hat R}_{00}-\frac 13\hat{\theta}^2,  
\eqno{(9)}
$$
where the overdot means derivative with respect to the transformed proper
time $\tau $ and $\hat{\theta}$ is the volume expansion. In eq.(9) we took
the reversed sense of time $-\infty \leq \tau \leq 0$, i.e. $\hat{a}$ runs
from infinity to zero. This equation can be finally written as:

$$
\hat{\theta}=-\frac 3{\hat{a}^2}-\frac{9\gamma M}{2\hat{a}^{3\gamma }}-\frac %
32\frac{(\omega +\frac 32)\;N}{\hat{a}^6}.  
\eqno{(10)}
$$

From it one sees that all terms in the right-hand side induce contraction
and hence a spacetime singularity is expected to occur (the global
singularity at $\hat a=0$).

The evolution of the volume expansion in the original frame, given in terms
of the conformal scale factor, can be easily found from eq.(10) if we
realize that $\theta =e^{\frac{{\psi }}2}(\hat{\theta}-%
\frac 32{\psi })$. We find that

$$
(\frac{d\theta }{dt})^{\pm }=3\frac{e^{{\psi }^{\pm }}}{%
\hat{a}^6}\{-\hat{a}^4-\frac 32\gamma M\hat{a}^{3(2-\gamma )}-(\frac \omega 2%
+1)N\pm \frac{\sqrt{N}}2\sqrt{\hat{a}^4+M\hat{a}^{3(2-\gamma )}+(\frac \omega
6\;+\frac 14)\;N}\},  
\eqno{(11)}
$$
where $t$ is the proper time in the original frame. It is related with $\tau 
$ through $dt=e^{-\frac 12{\psi }^{\pm }}d\tau $. The '+'
and '-' signs in eq.(11) correspond to two possible branches of the
solution. From (11) one sees that for the '+' branch of the solution, the
last term in brackets induces expansion.

We are interested now in the limiting case $\hat{a}\ll 1$, since the
singularity in the transformed frame is found at $\hat{a}=0$. In this case
for $\omega =-\frac 32$ eq.(11) can be written as:

$$
(\frac{d\theta }{dt})^{\pm }\approx \pm \frac{3\sqrt{NM}e^{\hat{\phi}_0}}{2%
\hat{a}^{\frac 32(\gamma +2)}}\exp [\mp \frac 23\sqrt{\frac NM}\frac{\hat{a}%
^{-\frac 32(2-\gamma )}}{(2-\gamma )}],  
\eqno{(12)}
$$
for $\gamma >\frac 23$. $\hat{\phi}_0$ is some integration constant. For $%
\gamma =\frac 23$ we obtain:

$$
(\frac{d\theta }{dt})^{\pm }\approx \pm \frac{3\sqrt{N(M+1)}e^{\hat{\phi}_0}%
}{2\hat{a}^4}\exp [\mp \frac 12\sqrt{\frac N{M+1}}\hat{a}^{-2}],  \eqno{(13)}
$$
while for $\gamma <\frac 23$;

$$
(\frac{d\theta }{dt})^{\pm }\approx \pm \frac{3\sqrt{N}e^{\hat{\phi}_0}}{2%
\hat{a}^4}\exp [\mp \frac{\sqrt{N}}2\hat{a}^{-2}].  
\eqno{(14)}
$$

For $\omega >-\frac 32$, in the limit $\hat{a}\ll 1$, eq.(11) can be written
in the following way:

$$
(\frac{d\theta }{dt})^{\pm }\approx \frac{N\;e^{\hat{\phi}_0}}{2\;\hat{a}%
^{6\mp \sqrt{\frac 6{\omega +\frac 32}}}}(-3\omega -\frac 92\pm 2\sqrt{%
6\omega +9}-\frac 32),  
\eqno{(15)}
$$
for $0<\gamma <2$.

A careful analysis of eq.(11) shows that, for big $\hat{a}$ the first three
terms in brackets prevail over the last one and, consequently, contraction
is favored until $\hat{a}$ becomes sufficiently small ($\hat{a}\ll 1$). In
this case, when $\omega $ is in the range $-\frac 32\leq \omega \leq -\frac 4%
3$, in the '+' branch of the solution, there are not enough conditions for
further contraction and the formation of the global singularity is not
allowed. A cosmological wormhole is obtained instead. For further
analysis of what happen in this case, we need to write down the relevant
magnitudes and relationships, in the original frame, in terms of the
transformed scale factor. We shall be interested in the behaviour of these
magnitudes and relationships for small $\hat{a}\ll 1$ (when the condition
for further contraction ceases to hold), in the '+' branch of the solution.
In this case the scale factor is found to be

$$
a^{+}\approx e^{-\frac 12\hat{\phi}_0}\hat{a}^{1-\frac 12\sqrt{\frac 6{%
\omega +\frac 32}}}.  
\eqno{(16)}
$$

The Ricci scalar is

$$
R^{+}\approx \frac 32Ne^{\hat{\phi}_0}\hat{a}^{\sqrt{\frac 6{\omega +\frac 32%
}}-6},  
\eqno{(17)}
$$
while for the proper time $t$ we have that

$$
t^{+}\approx \frac{2e^{-\frac 12\hat{\phi}_0}\hat{a}^{3-\frac 12\sqrt{\frac 6%
{\omega +\frac 32}}}}{\sqrt{\frac{(\omega +\frac 32)\;N}6}(6-\sqrt{\frac 6{%
\omega +\frac 32}})},  
\eqno{(18)}
$$
for $\omega \neq -\frac 43$ and

$$
t^{+}\approx \frac{e^{-\frac{1}{2}\hat{\phi}_{0}}}{\sqrt{\frac{(\omega +%
\frac{3}{2})\;N}{6}}}\ln \hat{a},  
\eqno{(19)}
$$
for $\omega =-\frac{4}{3}$. Hence, if we choose the '+' branch of the
solution, for $\omega \geq -\frac{4}{3}$, $R^{+}$ is bounded for $\hat{a}%
\rightarrow 0$. In this limit $a^{+}\rightarrow +\infty $ while $%
t^{+}\rightarrow -\infty $. A similar analysis shows that, for big $\hat{a}$%
, $\hat{a}\rightarrow \infty \Rightarrow a^{\pm }\rightarrow \infty $ and $%
t^{\pm }\rightarrow +\infty $. For intermediate values in the range $0<%
\hat{a}<\infty $ the curvature scalar $R^{+}$ is well behaved and bounded.
The scale factor $a$ is a minimum at some intermediate time $t_{\ast }$.
Hence in the original frame, if we choose the '+' branch of the solution,
the following picture takes place. If we restrict $\omega $ to fit into the
range $-\frac{3}{2}\leq \omega \leq -\frac{4}{3}$ and for $0<\gamma <2$,
then, the universe evolves from the infinite past $t=-\infty $ when he had
an infinite size, through a bounce at some intermediate $t_{\ast }$ when he
reached a minimum size $a_{\ast }$, into the infinite future $t=+\infty $
when he will reach again an infinite size. 

We now elaborate on $\psi $ evolution. Combining equations (6) and (16) , and
the fact that for $\hat{a}\ll 1$, we find that $\tau \sim \hat{a}^{3}$ in
the conformal frame, we obtain

$$
\psi =\psi _{0}+\frac{1}{3}\sqrt{\frac{6}{\omega +\frac{3}{2}}}\ln \tau 
\eqno{(20)}
$$

where $\psi _{0}$ is an arbitrary constant.

Extrapolating the validity of above equation to arbitrary times, we could
conclude that the dilaton (the curved extra-coordinate in our distorted RSL
scenario) evolves, in this case, from $\psi ^{+}\sim -\infty $ at $\tau
^{+}\sim 0$ into $\psi ^{+}\sim \infty $ at $t^{+}\sim +\infty $. As
illustrations to this behaviour we shall study the particular cases with $%
\omega =-\frac{3}{2}$ for dust-filled and radiation-filled universes since,
in these very particular situations exact analytic solutions can be easily
found.

For a radiation-filled universe ($\gamma =\frac 43$) the equation conformal
to (4), with $\omega =-\frac 32$ can be written as:

$$
\dot{\hat{a}}=\sqrt{M\hat{a}^{-2}+1},  
\eqno{(21)}
$$
and, after integration we obtain for the transformed scale factor

$$
\hat{a}=\sqrt{\tau ^2-M}. 
\eqno{(22)}
$$

The proper time $\tau $ is constrained to the range $\left| \tau \right|
\geq \sqrt{M}$ or $\sqrt{M}\leq \tau \leq +\infty $ (the case $-\infty \leq
\tau \leq -\sqrt{M}$ corresponds to the time reversed solution). The scale
factor, in our frame, is then found to be

$$
a^{\pm }=\frac{\sqrt{\tau ^2-M}}{\sqrt{\phi _0}}\exp [\pm \frac 12\frac{%
\sqrt{N}}M\frac \tau {\sqrt{\tau ^2-M}}],  
\eqno{(23)}
$$
while the curvature scalar:

$$
R^{\pm }=\frac{3}{2}N\phi _{0}\frac{\exp [\pm \frac{\sqrt{N}}{M}\frac{\tau }{%
\sqrt{\tau ^{2}-M}}]}{(\tau ^{2}-M)^{3}}.  
\eqno{(24)}
$$

The relationship between the proper time $\tau $ measured in the transformed
frame and the one in our initial frame (for the '+' branch of the solution
that is the case of interest) is given by

$$
t^{+}=-\frac \tau {\sqrt{M\phi _0}}\exp [\frac{\sqrt{N}}{2M}\frac \tau {%
\sqrt{\tau ^2-M}}].  \eqno{(25)}
$$

A careful analysis of eq.(23) shows that $a^{+}$ is a minimum at some time
that is a root of the algebraic equation $\tau ^{4}-M\tau ^{2}-\frac{N}{4}=0$%
. The curvature singularity occurring in the conformal frame at $\tau =\sqrt{%
M}$, is removed in our initial frame, where $R^{+}$ is bounded and well
behaved for all times in the range $\sqrt{M}\leq \tau \leq +\infty $ ($%
-\infty \leq t\leq +\infty $).

\bigskip Combining equations (6) and (23) we get

$$
\psi =\psi _{0}-\frac{\sqrt{N}}{M}\frac{\tau }{\sqrt{\tau ^{2}-M}}.
\eqno{(26)}
$$

We see that for $\tau \rightarrow \sqrt{M},\psi \rightarrow -\infty $ and
for\ $\tau \rightarrow \infty ,\psi \rightarrow \psi _{0},$ so the
''visible'' brane starts from an infinite negative separation and
asymptotically  tends to a finite separation at the infinite future. This
way, brane stabilization occurs. In this case conformal (Weyl) symmetry is
continously broken and standard general relativity over Riemann geometry is
recovered asymptotically in the future.

For a dust-filled universe ($\gamma =1$) the transformed scale factor can be
given the form:

$$
\hat{a}=\frac{4M}{\eta ^2-4},  \eqno{(27)}
$$
where the time variable $\eta $ has been introduced through $d\tau =\frac{%
\hat{a}^2}Md\eta $ and is constrained to the range $2\leq \eta \leq +\infty $
(the case $-\infty \leq \eta \leq -2$ corresponds to the time reversed
solution). In our initial frame we have that

$$
a^{\pm }(\eta )=\frac{4M}{\sqrt{\phi _0}}\frac{\exp [\mp \frac{\sqrt{N}}{%
24M^2}\eta (\eta ^2-12)]}{\eta ^2-4},  \eqno{(28)}
$$
and the relationship between the proper time $t$ and $\eta $ is given by the
following expression:

$$
t^{\pm }=\frac{16M}{\sqrt{\phi _{0}}}\int d\eta \frac{\exp [\mp \frac{\sqrt{N%
}}{24M^{2}}\eta (\eta ^{2}-12)]}{(\eta ^{2}-4)^{2}}.  \eqno{(29)}
$$

The curvature singularity occurring in the conformal frame at time $\eta =2$
is removed again in our initial representation of the theory. The '+' branch
scale factor $a^{+}$ is a minimum at some $\eta _{\ast }$ that is a root of
the algebraic equation $\eta ^{4}-8\eta ^{2}+\frac{16M^{2}}{\sqrt{N}}\eta
+16=0$.

\bigskip For the dilaton we have

$$
\psi =\psi _{0}+\frac{\sqrt{N}}{12M^{2}}\eta (\eta ^{2}-12)
\eqno{(30)}
$$

We see that for $\eta \rightarrow 2,\psi \rightarrow \psi _{0}-\frac{4\sqrt{N%
}}{3M^{2}}$ and for\ $\eta \rightarrow \infty ,\psi \rightarrow \infty ,$ so
the ''visible'' brane starts from an arbitrary finite negative
separation and  tends to an infinite separation at the infinite future. This
way, brane unstabilizes and conformal (Weyl) symmetry is recovered in the
future. 

Summing up, from all above results we could conclude that in this 
scenario an open FRW evolves from a radiation dominated universe in a
visible brane infinitely separated (in the negative $\psi $ direction) from
the Planck brane to a dust universe again infinitely separated from the
reference (Planck) brane, but now in the positive $\psi $ direction, while
in reference \cite{qbc}, the flat universe goes from an infinite separation
in the distant past to a finite separation in the future. So, for open
universes no brane stabilization occurs, and a more dynamic picture
respect to the results obtained for flat universes takes place. In this case
conformal (Weyl) symmetry was continously broken and standard general
relativity over Riemann geometry was recovered asymptotically, while the
inverse picture will occur in the future. Above results also tell us that
both flat and open universes, are free of the cosmological singularity in
our initial formulation, in the same region of the parameter space ($-\frac{4%
}{3}\leq \omega \leq -\frac{3}{2}$, $0<\gamma <2$). The fact that our
non-singular branch is the '+' one, instead of the '-' branch in reference 
\cite{qbc} is simply due to the change of the sign before $\psi $ in the
exponents of our action.

Finally we shall remark the fact that the cosmological singularity is
removed, in our initial frame, only for a given range of the parameter $%
\omega $. It can be taken just as a restriction on the values this parameter
can take. A physical consideration why we chose the '+' branch (the
non-singular branch) instead of the '-' branch is based on the following
analysis. We shall note that in our frame, $e^{-{\psi}}$ plays the role
of an effective gravitational constant $G$. For the '-' branch $G$ runs from
zero to an infinite value, i.e. gravity becomes stronger as the universe
evolves and, in the infinite future it dominates over the other
interactions, that is in contradiction with the usual picture. On the
contrary, for the '+' branch, $G$ runs from an infinite value to zero and
hence gravitational effects are weakened as the universe evolves, as
required.

\end{document}